\documentclass[aps, prb, twocolumn, showpacs, floatfix,10pt,superscriptaddress]{revtex4-2}
\usepackage{amsmath}
\usepackage{amsfonts}
\usepackage{amsbsy}
\usepackage{amssymb}
\usepackage{graphicx}
\usepackage{textcomp}
\usepackage[caption=false,singlelinecheck=false]{subfig}
\usepackage{xcolor}
\usepackage{mathrsfs}
\usepackage{mathtools}
\usepackage{bm}
\usepackage{gensymb}
\usepackage{braket,wasysym}
\usepackage[colorlinks,linkcolor=blue,citecolor=red,filecolor=magenta,urlcolor=red,breaklinks]{hyperref}
\usepackage[bottom]{footmisc}
\usepackage{verbatim}

\hypersetup{colorlinks=true, urlcolor=blue, citecolor=magenta, pdfborder={0 0 0}}
\usepackage{breakurl}
\usepackage{natbib}

\allowdisplaybreaks

\normalfont
\def\be{\begin{equation}}
	\def\ee{\end{equation}}
\def\bea{\begin{eqnarray}}
	\def\eea{\end{eqnarray}}

\begin{document}
\title{Pattern-dependent proximity effect and Majorana edge mode
\\in one-dimensional quasicrystals}

\author{Junmo Jeon}
\email{junmo1996@kaist.ac.kr}
\affiliation{Korea Advanced Institute of Science and  Technology, Daejeon 34141, South Korea}
\author{SungBin Lee}
\email{sungbin@kaist.ac.kr}
\affiliation{Korea Advanced Institute of Science and  Technology, Daejeon 34141, South Korea}

\date{\today}
\begin{abstract}
The Majorana edge states of the Kitaev chain model have attracted extensive attention on their stability and experimental realization. One of the interesting aspects is finding the exotic proximity effect, which guarantees the presence of the Majorana modes, further enables efficient braidings between them.
In this paper, we explore the superconducting proximity effect for quasi-periodic quantum wires and discuss how quasi-periodic patterns affect the stability of the Majorana modes. 
Considering the Kitaev chain model of the one-dimensional quasi-periodic system, we discuss the pattern-dependent proximity effects. First, we argue that the presence of quasi-periodic hoppings energetically induces the $p$-wave pairing also to be quasi-periodic rather than uniform pairing.
More interestingly, when the normal metallic wire is adjacent to the quasiperiodic superconducting wire, we have found that the Majorana edge modes are being transferred to the edge of the normal metallic side with enhanced stability.
Finally, we discover the proximity effect on the strengths of the quasi-periodicities with a general power-law relationship, whose power depends on the tiling pattern.
Our results show how quasi-periodic patterns play an important role in the Kitaev chain and the stabilization of the Majorana mode.  
\end{abstract}
\maketitle

\section{Introduction}
\label{sec:introduction}
Over a decade, the topological Majorana edge mode, theoretically predicted by the Kitaev chain model, 
is one of the most studied physical phenomena in condensed matter physics\cite{kitaev2001unpaired,greiter20141d,wolms2012majorana,cook2012stability,dumitrescu2013topological,leumer2021linear,bernevig2013topological,altomare2013one}. The topological robustness of the Majorana edge mode has been attracted broad research interest such as information theory and quantum computing. Also, it has been discussed in experimental realization with general aspects. \cite{van2016information,levy2019entanglement,rieffel2011quantum,nielsen2002quantum,huang2021emulating,aaronson2013quantum,xiao2020topological,sau2012realizing,tutschku2020majorana,lian2018topological,gangadharaiah2011majorana,degottardi2013majorana,sengupta2009entanglement,PhysRevResearch.3.013148}.
Along with it, the stability of the Majorana edge modes has been studied in the presence of spatially varying hopping amplitudes or inhomogeneous potentials, for instance.
One of the interesting questions is how to realize the Majorana modes and the proximity effect when the system is neither periodic nor random. Recently, there are relevant studies for quasi-periodically driven systems to look for possible topological phases and their robust edge modes\cite{crowley2019topological,friedman2020topological,ray2019dynamics}.
Nevertheless, for a quasi-periodic system with spatial complexity, their existence and the pattern-dependent proximity effect have been barely understood
\cite{ghadimi2020topological,sakai2019exotic,roche1997electronic,ghadimi2017majorana}.

To demonstrate the braiding operation of the Majorana modes, their stability is the most important issue to prevent thermalization \cite{xiao2020topological,tutschku2020majorana,rieffel2011quantum}.
Since the stability of the Majorana edge mode is related to the size of the topological gap, the energy spectrum of the entire system itself should be investigated to look for the stable Majorana mode. In this vein, one would offer the quasi-periodic system as an opportunity to realize a more stable Majorana mode due to its unconventional multi-fractal energy spectrum. Generally, it is known that there are multi-fractal sizes and positions of the gaps in the quasi-periodic systems due to their exotic tiling patterns\cite{jeon2021topological,jeon2020phonon,kellendonk2015mathematics,sadun2008topology,bellissard1992gap}. Thus, one can speculate that the quasi-periodic system could enlarge the gap size and guarantee enhanced stability of Majorana mode that is strongly dependent on their unique tiling patterns.

\begin{figure}[]
	\centering
	\includegraphics[width=0.45\textwidth]{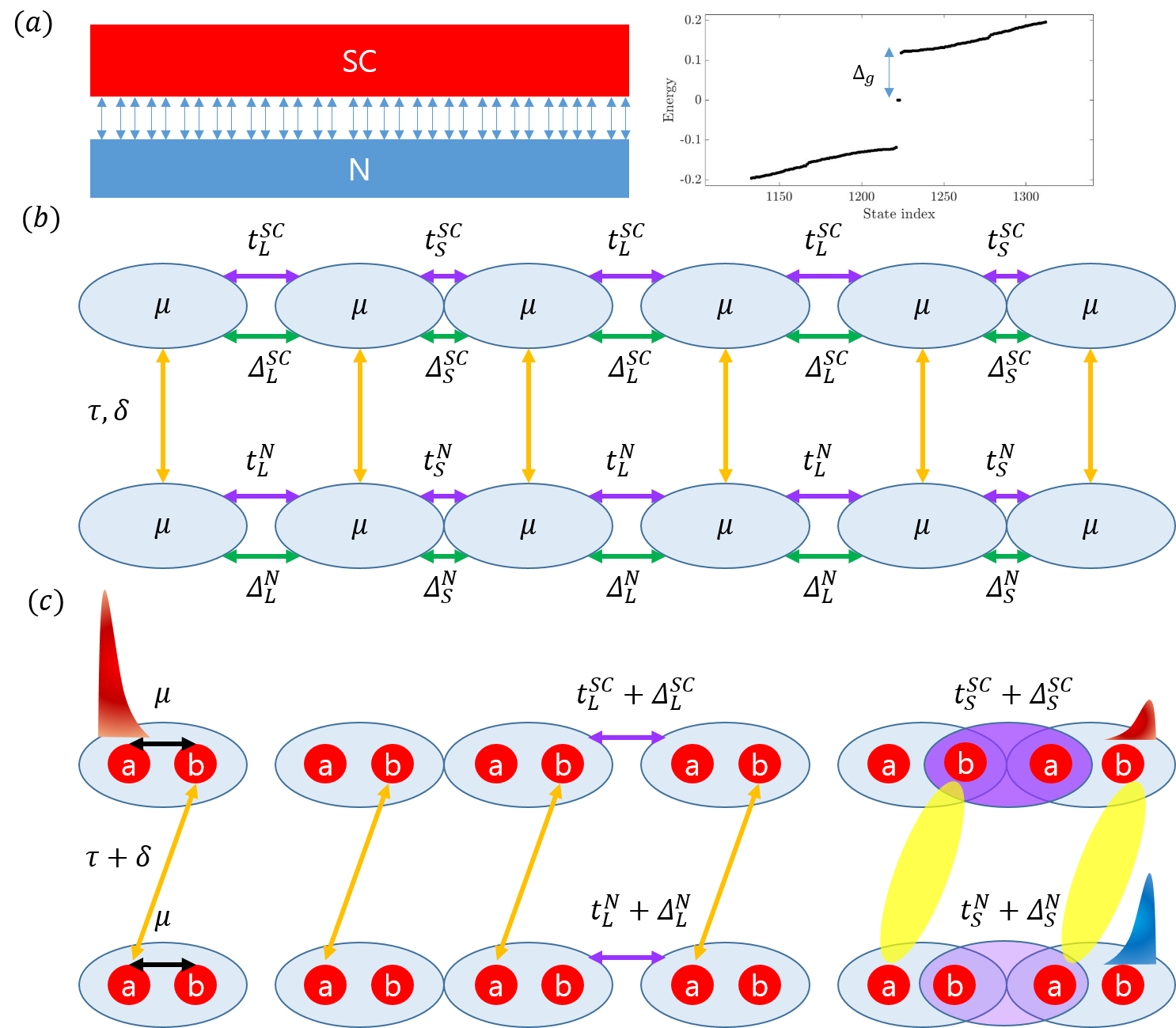}
	\label{fig3}
	\caption{\label{fig:summary} Schematic illustration of the proximity induced Majorana edge mode in the stacked quantum wires. (a) Stacking of  the superconducting quantum wire (SC) on the trivial normal metal (N) and its typical energy spectrum. The size of topological gap $\Delta_g$ is defined as in the right panel. (b)  The hopping magnitudes and pairing potentials are arranged in a quasi-periodic way with a chemical potential $\mu$. Super scripts (SC/N) for a superconducting wire and a normal metal. Subscript ($L/S$) for two kinds of prototiles. $t$ and $\Delta$ stand for hoppings and pairings in each wire. The proximity effect comes from the inter-wires hoppings and pairing potentials, given by $\tau$ and $\delta$, respectively. 
	 (c) Given $\tau=\delta$ limit, the coupling represented by the dotted arrow is negligible. In contrast, the $a$-Majorana fermions of the normal metal couple to the $b$-Majorana fermions of the superconductor due to the inter-wires hoppings and pairings.  As a result, the proximity $b$-Majorana edge mode emerges (blue envelope). See the main text for details.
	}
\end{figure}

Focusing on the pattern-dependent proximity effect in quasi-periodic superconducting wires, our main results are summarized as following: 
(i) For the quasi-periodic wire where the quasi-periodicity is encoded in hopping magnitudes, the system prefers the superconducting pairing also to be quasi-periodic, thus the system generically realizes quasi-periodicity in both hoppings and pairings. (ii) When the normal metallic wire is placed on bottom of the quasi-periodic superconducting wire (See Fig.\ref{fig:summary} (a)), the proximity effect induces normal metal  having the same tiling pattern as the quasi-periodic wire and  the Majorana edge modes is being transferred to the normal metallic wire with enhanced stability (See Fig.\ref{fig:summary} (c)).
   Furthermore, it turns out that the  strength of the induced quasi-periodicity in normal metal, $\rho^N_*$, is proportional to the one for quasiperiodic superconducting wire showing a power-law relationship,
\begin{align}
\label{mainresult}
 \rho^{N}_*=(\rho^{SC})^\beta,
\end{align}
where $\rho^{SC(N)}_{(*)}=t_L^{SC(N)}/t_S^{SC(N)}$ are the hopping ratios considering the quasiperiodicity composed of the two hopping magnitudes, say $t_L$ and $t_S$, in the superconductor (SC) and the normal metal (N), respectively. Here, the power, $\beta>0$, depends on the specific tiling pattern. With the relation in Eq.\eqref{mainresult}, the Majorana edge mode shows enhanced stability in quasi-periodic systems than uniformly periodic cases. In addition, the influence of the disorder is also discussed. Our results suggest potential topological quantum computing in the quasi-periodic systems with more stable Majorana modes.

%organization
This article is organized as following. In Sec.\ref{sec:pairing}, we introduce the Kitaev chain model and self-consistently calculate local pairing potential when the hopping terms exhibit quasi-periodic ordering. In Sec.\ref{sec:stacking}, we consider the stacked quantum wires of superconductor and normal metal and discuss the stability of the proximity induced Majorana edge mode on the normal metal. We discover the interesting power-law relationship between the strengths of the quasi-periodicities in each wire, that is uniquely determined by the quasi-periodic tiling patterns. Also, we demonstrate the proximity induced Majorana edge mode has enhanced stability with larger gap size for the quasi-periodic system rather than uniformly periodic one.
Considering several tiling patterns, we argue that such proximity effects and enhanced stability of Majorana edge mode are indeed present in many quasi-periodic systems.
Finally, we conclude our work in Sec.\ref{sec:conclusion}. 
%This article is organized as following. In Sec.\ref{sec:pairing}, we introduce the Kitaev chain model and self-consistently calculate local pairing potential when the hopping terms exhibit quasi-periodic ordering. Based on the self-consistent pairing potentials, we demonstrate the proximity effect on the pairing potential due to the quasi-periodic hoppings. In Sec.\ref{sec:stacking}, we introduce the stacked quantum wires (superconductor and normal metal, respectively) model and discuss the stability of the proximity induced Majorana edge mode on the normal metal, which is given by the size of the topological gap. We discover the interesting power-law relationship between the strengths of the quasi-periodicities in each wire as Eq.\eqref{mainresult}. Also, we demonstrate the proximity induced Majorana edge mode has enhaced stability as larger size of the gap for the quasi-periodic system rather than uniformly periodic one. We examine that these proximity effects are indeed pattern-dependent one by considering various tiling patterns. We make the conclusion from our works in Sec.\ref{sec:conclusion}.

\section{Proximity effect on the pairing potentials}
\label{sec:pairing}
To study the pattern-dependent proximity effect, we start with introducing the microscopic model as following.
\begin{align}
\label{microscopic}
\mathcal{H}=\sum_{i=1}^{M}-t_{i,i+1}c_{i+1}^{\dagger}c_i+h.c.+\mu\sum_{i=1}^{M+1}c_i^{\dagger}c_i+\sum_{i=1}^{M}U_{i,i+1}n_in_{i+1}
\end{align}
Here, the first term is hopping terms representing kinetic energy part where $c_i(c_i^{\dagger})$ is fermionic annihilation (creation) operator on the $i$-th site. $t_{i,i+1}$ is hopping magnitude in between $i$-th site and $(i+1)$-th site. The second term represents the on-site energy and $\mu$ is a chemical potential. The number of sites is $M+1$, with $M$ number of links. The last term represents nearest neighbor interaction which is responsible for the pairing potential. Here, $U_{i,i+1}$ is the strength of the interaction in between $i$-th site and $(i+1)$-th site. In this paper, we consider $U_{i,i+1}<0$ i.e. attraction potential. Then, within the mean field decoupling, one can get the simple Kitaev chain model given by,
\begin{align}
\label{hamiltonian}
\mathcal{H}= \sum_{i=1}^{M}\left(-t_{i,i+1}c_{i+1}^{\dagger}c_i +\Delta_{i,i+1}c_ic_{i+1}+h.c.\right)+\mu\sum_{i=1}^{M+1}c_i^{\dagger}c_i.
\end{align}
%Here, $c_i(c_i^{\dagger})$ is fermionic annihilation (creation) operator on the $i$-th site. $t_{i,i+1}$ and 
Here, $\Delta_{i,i+1}$ is pairing potential in between $i$-th site and $(i+1)$-th site which is determined by the Bogoliubov-de-Gennes (BdG) Hamiltonian and the self-consistent equation\cite{cohen2016fundamentals,araujo2019conventional},
\begin{align}
\label{pairing potential}
\Delta_{i,i+1}=U_{i,i+1}\langle c_{i+1}^{\dagger}c_i^{\dagger}\rangle
\end{align}
Note that $t_{i,i+1},U_{i,i+1},\Delta_{i,i+1}$ and $\mu$ are considered in the energy unit throughout the paper. One can constructthe real space BdG Hamiltonian \cite{gomez2018universal}. In terms of quasi-periodic chemical potential, there are studies related to the topological phase transition\cite{degottardi2013majorana,ghadimi2017majorana}. In our study, we focus on quasi-periodicity encoded in the hopping magnitudes and pairing potentials in terms of $t_{i,i+1}$, $U_{i,i+1}$, and hence $\Delta_{i,i+1}$, keeping a constant chemical potential, $\mu$. In this case, we note that our model is exactly mapped onto the XY spin chain with quasi-periodic spin exchanges under the uniform external magnetic field along $z$-direction,
\begin{align}
\label{XYmodel}
\mathcal{H}_{XY}=\sum_{i=1}^{M}\left(J_i^{x}S_i^xS_{i+1}^x+J_i^{y}S_i^yS_{i+1}^y\right)+h\sum_{i=1}^{M+1}S_i^z
\end{align}
In Eq.\eqref{XYmodel}, $J_i^{x},J_i^y$ are spin exchange energies in between $i,i+1$ sites for $x,y$ spin, respectively. $h$ is the unifrom magnetic field. In details, $\frac{J_i^{x}+J_i^{y}}{4}=t_{i,i+1}$, $\frac{J_i^{x}-J_i^{y}}{4}=\Delta_{i,i+1}$ and $\mu=h$ under the Jordan-Wigner transformation\cite{parkinson2010introduction}. Thus, one can experimentally control both chemical potential and pairings in the Kitaev chain model in terms of the magnetic field strength and anisotropy of $J_i^x$ and $J_i^y$ in the anisotropic XY spin chain system.\\

From the Kitaev chain model in Eq.\eqref{hamiltonian}, the Majorana fermion operators, $a_j,b_j$ are defined by $c_j=\frac{1}{2}(a_j+ib_j)$, and obey anti-commutation relations; $\{a_i,b_j\}=0, \{a_i,a_j\}=2\delta_{i,j}=\{b_i,b_j\}$.
In terms of the $a_j,b_j$, we can rewrite the Hamiltonian in Eq.\eqref{hamiltonian} in $2(M+1)$-dimensional Hilbert space as Eq.\eqref{hamiltonianMajorana},
\begin{widetext}
\begin{align}
\label{hamiltonianMajorana}
\to\mathcal{H}-\frac{\mu}{2}=\frac{i}{2}\sum_{i=1}^{M}\left((-t_{i,i+1}-\Delta_{i,i+1})a_{i+1}b_i+(t_{i,i+1}+\Delta_{i,i+1})a_ib_{i+1}\right)+i\frac{\mu}{2}\sum_{i=1}^{M+1}a_ib_i.
\end{align}
\end{widetext}
The Hamiltonian in Eq.\eqref{hamiltonianMajorana} is hermitian because of the anti-commutation relationship of the Majorana operators. Then, the Majorana edge mode exists if the Majorana charge operators, $Q_a=\sum_{i=1}^M\alpha_ia_i$ and $Q_b=\sum_{i=1}^M\beta_ib_i$ commute with $\mathcal{H}$, where $\alpha_i$, $\beta_i$ are real coefficients\cite{kitaev2001unpaired}. Hence, the coefficients $\alpha_i,\beta_i$ should obey the recursion relationships, which can be written in terms of the transfer matrices, $M_i$ and $N_i$ as following. 
\begin{subequations}
\label{conditiontransfer}
\begin{equation}
\label{conditiontransfer1}
\begin{pmatrix} \alpha_{i+1} \\ \alpha_i \end{pmatrix}\!=\!M_i \begin{pmatrix} \alpha_i \\ \alpha_{i-1} \end{pmatrix}\!, ~M_i \!\equiv\!\begin{pmatrix} \frac{-\mu}{\Delta_{i,i+1}-t_{i,i+1}} & \frac{\Delta_{i-1,i}+t_{i-1,i}}{\Delta_{i,i+1}-t_{i,i+1}} \\ 1 & 0 \end{pmatrix}\!,
\end{equation}
\begin{equation}
\label{conditiontransfer2}
\begin{pmatrix} \beta_{i+1} \\ \beta_i \end{pmatrix}\!=\!N_i\begin{pmatrix} \beta_i \\ \beta_{i-1} \end{pmatrix}\!,~ N_i \!\equiv\!\begin{pmatrix} \frac{\mu}{\Delta_{i,i+1}+t_{i,i+1}} & \frac{\Delta_{i-1,i}-t_{i-1,i}}{\Delta_{i,i+1}+t_{i,i+1}} \\ 1 & 0 \end{pmatrix}.
\end{equation}
\end{subequations}
Then, the global transfer matrix is defined by $\mathcal{A}_p^{a(b)}\equiv\prod_{i=1}^M M_i$ ($\prod_{i=1}^M N_i$) for $a_j$ ($b_j$) Majorana fermions. With the global transfer matrix, $\mathcal{A}_p^{a(b)}$, the normalization condition is fulfilled when the topological invariant $\nu\equiv-\mbox{sgn}(f(1)f(-1))=-1 \mbox{ where } f(z)=\mbox{det}(\mathcal{A}^{a(b)}_p-zI)$\cite{degottardi2013majorana,degottardi2011topological}. 
Here, $\mbox{sgn}(x)$ is the sign of the real number $x$. See Supplementary Materials Sec.1 for the details. 

To discuss the pattern-dependent proximity effect, we first exemplify the case for the Fibonacci quasicrystal.  A Fibonacci quasicrystal is comprised of two prototiles that are termed by long ($L$) and short ($S$), and is generated by the iterative substitutions, $L\to LS$ and $S\to L$\cite{kellendonk2015mathematics,jeon2021topological,kohmoto1987electronic,gumbs1988dynamical,fang2016unexpected}. The resultant pattern becomes $LSLLSLSLLSLLS\cdots$.
Then, the nearest neighbor interaction and hopping magnitudes are assigned as $U_{L/S},t_{L/S}$ for each $L$ and $S$ tile, respectively. In terms of the hopping magnitude, the strength of the quasi-periodicity is given by $\rho\equiv t_L/t_S$. 

Now we investigate the proximity effect on electron pairings induced by the quasi-periodic hoppings, $t_L$, $t_S$ and their ratio $\rho$. First of all, since we keep $\mu$  as a constant, the physical meaning of $\rho$ is related to the relative distance between the sites in the quantum wire. Specifically, $\rho<1 (>1)$ tells us that the distance of $L$ link is longer (shorter) than $S$ link.
Thus, it is reasonable to consider $U_L/U_S< 1$ for $\rho< 1 $ and $U_L/U_S> 1$ for $\rho> 1 $, respectively.  For the periodic case, $U_L=U_S$ when $\rho=1$. Now, with a given value of $\rho$, we self-consistently compute $\Delta_L,\Delta_S$ under the microscopic interaction strengths $U_L,U_S$. 
{Here, $\Delta_L,\Delta_S$ are defined by the mean value of the self consistent pairings for $L$,$S$ links, respectively.} The system size is $M=F_{15}=610$, where $F_n$ is the $n$-th Fibonacci number. To ensure the presence of Majorana edge mode, we have set $\mu=0$ (See Eq.\eqref{quasiv}.). As a result, if $\rho<1 (>1)$, then $\kappa\equiv\Delta_L/\Delta_S<1 (>1)$ in whole physical parameters $U_L,U_S$. Thus, It indicates that the proximity effect of the quasi-periodic hoppings gives rise to the pairings which follow the same quasi-periodic pattern.

Based on the proximity effect of the quasi-periodic hoppings, now one can determine the regime where the topological phase is stabilized in the Fibonacci chain with the quasi-periodic hoppings and pairings. 
%to ensure each wire in Fig.\ref{fig:summary} as topological superconductor and trivial normal metal, respectively. 
In the limit $\Delta_{i,i+1}=t_{i,i+1}$, one can simply derive such condition based on Eq.\eqref{conditiontransfer}   
%Since now the second column of the matrix Eq.\ref{conditiontransfer} vanishes, their iterative product $\mathcal{A}_p$ also has zero second column. Thus,
and the topological invariant $\nu$ becomes
\begin{align}
\label{quasiv}
\nu=\mbox{sgn}\left(\left(\frac{\mu}{2t_S\rho^{\alpha(M)}}\right)^{2M}-1\right).
\end{align}
In Eq.\eqref{quasiv}, the topological phase is given by $\nu=-1$ when $\left\vert \mu\right\vert<\left\vert2t_S\rho^{\alpha(M)}\right\vert$.
Here, $\alpha(M)$ is a relative portion of the $L$ tiles among the total $M$ number of tiles. In the Fibonacci case, for instance, $\alpha(M)$ saturates to the golden ratio in the thermodynamic limit $M\rightarrow \infty$. 
Hence, depending on the quasi-periodic patterns, the condition for the topological phase changes. See Supplementary Materials Sec.2 for detailed information.
%\tcr{so what? mention what is the consequence of alpha(M) approaches to the golden ratio}
%\tcr{no explanation of Fig.2(b)} 

%\tcg{From Eq.\eqref{quasiv}, we achieve the set up for the pattern-based proximity effect in between two wires in Fig.\ref{fig:summary} by assigning the larger hopping magnitudes on the superconducting wire than the trivial normal metal.}

\section{Proximity effect of stacked quantum wires} 
\label{sec:stacking}
Given that the quasi-periodic tiling pattern plays an important role in pairing formation, now we ask how such pattern complexity stabilizes the Majorana edge mode and the topological gap is enhanced. Although the robustness of the Majorana edge mode with respect to the disorder has been explored in many literatures\cite{zhang2019majorana,degottardi2013majorana}, its stability with respect to the thermalization which is characterized by the size of the topological gap, is barely understood. However, such inquiry is genuinely crucial because the excitation of the Majorana mode is a critical obstacle in topological quantum computing. Hence, to prevent thermalization, the larger gap size is more efficient.
Moreover, since the spectral gap strongly depends on the quasi-periodic tiling pattern in the bulk\cite{jeon2021topological,jeon2020phonon,kellendonk2015mathematics,sadun2008topology,bellissard1992gap}, the stability issue of the Majorana edge mode itself is a pattern-dependent proximity effect. For stable operation of the Majorana modes, therefore, it is important to study the stability of the Majorana edge mode as function of the tiling patterns.

In this vein, we ask how quasi-periodic tiling pattern of the quantum wire affects to the stability of the Majorana mode and its proximity effect when the superconducting wire is stacked on top of the normal metal, as shown in Fig.\ref{fig:summary} (a). Specifically, stacking of two quasi-periodic wires, one for topological superconductor and another for trivial metal, is considered. Now we turn on the inter-layer hoppings ($\tau$) and pairings ($\delta$) between the wires. Then, the proximity induced Majorana edge mode is formed at the end of the trivial metal (illustrated as blue envelope in Fig.\ref{fig:summary}(c)). See Supplementary Materials Sec.3 for detailed information of the proximity induced Majorana edge mode.
%%%%%%additional part
Now, the Hamiltonian is composed of three parts as $\mathcal{H}=\mathcal{H}_{SC}+\mathcal{H}_N+\mathcal{H}_{interwires}$. Each tern is given by,
\begin{align}
&\mathcal{H}_{SC}= \sum_{i=1}^{M}\left(-t_{i,i+1}^{SC}c_{i+1}^{\dagger}c_i +\Delta_{i,i+1}^{SC}c_ic_{i+1}+h.c.\right)+\mu\sum_{i=1}^{M+1}c_i^{\dagger}c_i, \\
&\mathcal{H}_{N}= \sum_{i=1}^{M}\left(-t_{i,i+1}^{N}c_{i+1}^{\dagger}c_i +\Delta_{i,i+1}^{N}c_ic_{i+1}+h.c.\right)+\mu\sum_{i=1}^{M+1}c_i^{\dagger}c_i, \\
\label{interlayerH}
&\mathcal{H}_{interlayer}= \sum_{i=1}^{M+1}\left(-\tau c_{i(SC)}^{\dagger}c_{i(N)} +\delta c_{i(SC)}c_{i(N)}+h.c.\right).
\end{align}
Here, $t^{SC/N}$ and $\Delta^{SC/N}$ in $\mathcal{H}_{SC/N}$ represent the hopping magnitude and pairing potential in the superconductor and normal qauntum wire, repsectively. In Eq.\eqref{interlayerH}, $i(SC),i(N)$ are the $i$-th site indices in the superconductor and normal metal, respectively. Given length $M$ quantum wires, there are $2(M+1)$ number of Majorana fermions in each wire. Thus, we have $4(M+1)$ by $4(M+1)$ size Hamiltonian matrix for the stacked quantum wires system  (See Fig.\ref{fig:summary} (c) for $M=5$ example with 24 Majorana fermions.).

\begin{figure}[]
	\centering
	\includegraphics[width=0.5\textwidth]{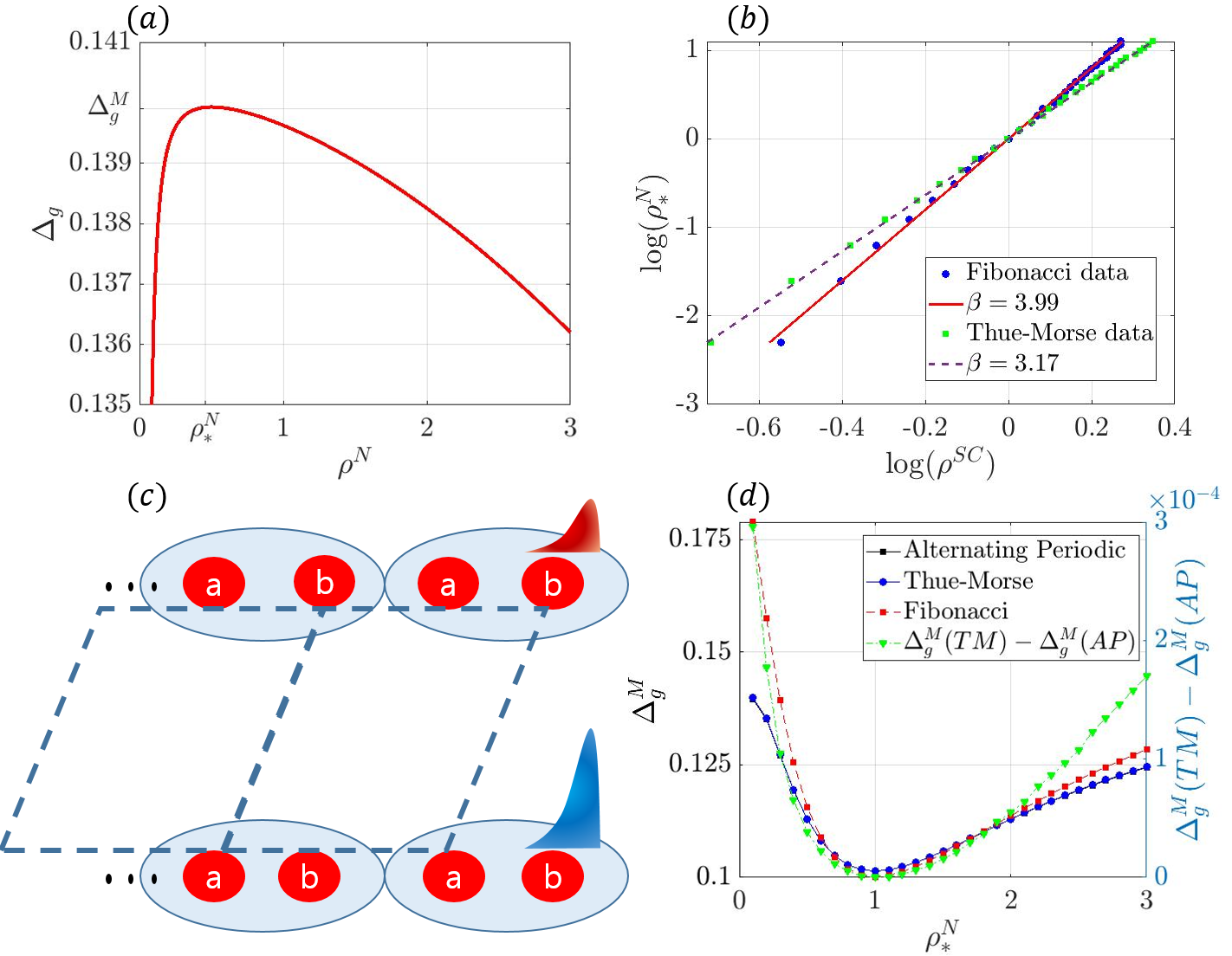}
	\label{fig1}
	
	\caption{\label{fig:result} Pattern-dependent stability of Majorana mode in the stacked wires.
(a) General concave behavior of the topological gap, $\Delta_g$, as a function of induced quasi-periodicity of the normal metal, $\rho^{N}$. For a given $\rho^{SC}$, there is a maximum size of the topological gap, $\Delta_g^M$ when $\rho^N=\rho^N_*$. (b) The power-law relationship between optimal strengths of the induced quasi-periodicity in the trivial metal, $\rho_*^N$, and the quasi-periodicity in the superconducting wire, $\rho^{SC}$. The power $\beta$ depends on the tiling pattern. (c) A schematic picture of a parallelogram cell formed by Majorana fermions. Stability of the Majorana edge modes is originated from the robustness of the pairings along these parallelogram cells. (d) Maximal topological gap, $\Delta_g^M$, as a function of $\rho_*^N$ for Fibonacci (red squares) and Thue-Morse (blue circles) cases. The value of $\Delta_g^M$ is minimum in the limit of periodic case, $\rho^N=\rho^{SC}=1$. 
See the main text for details.
	}
\end{figure}

Although the formation of such proximity induced Majorana edge mode is generic in stacked wire, regardless of the specific tiling pattern, how strongly it is stabilized depending on the tiling pattern is an important issue. %In this vein, the tiling pattern would play the significant role in the stacked wires system.
Considering $\tau=\delta$, the paired Majorana fermions form the parallelogram cells sharing the edge whose verticies are $b$-, and $a$-Majorana fermions in superconductor and trivial metal, respectively (See dashed parallelograms in Fig.\ref{fig:result} (c).). By forming the parallelogram cells, they are not independent but affect to each others via the shared edge depending on the given tiling pattern. Furthermore, since the proximity induced Majorana edge mode is more stable when such pattern-dependent pairings become stronger, the same complex pattern of the superconducting wire is also induced to the normal metal which we discuss below. %\tcg{Since the stability of the proximity induced Majorana edge mode originates from the robustness of \tcg{such parallelogram pairings} \tcr{wording}, one can understand the stability of the proximity induced Majorana edge mode through the tiling pattern in the stacked wires system.}
%\tcr{better to rewrite }

%To address such pattern-dependent proximity effect, 

To explore how the proximity effect stabilizes the Majorana edge mode in the stacked quasi-periodic wires, we survey the size of the topological gap, $\Delta_g$, depending on the tiling pattern and the strength of the quasi-periodicity, $\rho^{(SC/N)}\equiv t^{(SC/N)}_L/t^{(SC/N)}_S$ in superconductor (SC) and trivial normal metal (N). Here, the size of the topological gap, $\Delta_g$ is defined as shown in Fig.\ref{fig:summary} (a).
We consider the limit, $\Delta_{i,i+1}=t_{i,i+1}$ and $\tau=\delta$, keeping $\kappa^{(SC/N)}=\rho^{(SC/N)}$.
Since the gap size depends on the average pairing potentials, we should keep them as $c^{(SC/N)}$ constants.
Note that $c^{(SC/N)}={\sum_{i=1}^{M}\Delta_{i,i+1}^{(SC/N)}}/{M}$.  Based on Eq. \eqref{quasiv}, to distinguish the superconductor and the normal metal, we have $c^{(SC)}>c^{(N)}$. For instance, the Fibonacci quasicrystal with a length $M=F_n$, where $F_n$ is the $n$-th Fibonacci number, has $F_{n-1}$ and $F_{n-2}$ number of long and short tiles respectively. Thus, for given $\rho^{(SC/N)}$ we set 
\begin{align}
\label{parametersetting}
\Delta_S^{(SC/N)}=t_S^{(SC/N)}=\frac{c^{(SC/N)}F_n}{F_{n-2}+\rho^{(SC/N)}F_{n-1}}
\end{align}
and $\Delta_L^{(SC/N)}=t_L^{(SC/N)}=\rho^{(SC/N)}t_S^{(SC/N)}$. For numerical computation, we have set $c^{(SC)}=10$ and $c^{(N)}=0.8$ for $\mu=2$,  $\tau=\delta=1$ and $M=F_{15}=610$. Note that from Eq.\eqref{XYmodel}, our model is equivalent to the ising spin model ($J_i^y=0$) under the uniform magnetic field.

By exploring the proximity induced Majorana edge mode in Fibonacci pattern, we figure out two important characters.
One is that the pattern complexity in the normal metal prefers to follow the same pattern of the superconducting wire. In this case, the strengths of the quasi-periodicities have the power-law relationship as shown in Fig.\ref{fig:result}{(b)}. 
Specifically, there is an optimized value of $\rho^N$, say $\rho_*^N$, for $\Delta_g^M$ which maximizes the gap size $\Delta_g$ (See Fig.\ref{fig:result} (a).). Such optimized value $\rho_*^N$ is proportional to the strength of the quasi-periodicity in the superconducting wire, $\rho_*^N=(\rho^{SC})^\beta$  (See Fig.\ref{fig:result}(b).). Here, the positive value of $\beta$ indicates the proximity induced pattern in the normal metal indeed follows the original tiling pattern of the superconducting wire. The magnitude of $\beta$ depends not only on the quasi-periodic pattern but also on the strengths of the inter-wires hoppings and pairings. In particular, for given parameters $\tau=\delta=1$, we get $\beta\approx3.99$ shown in Fig.\ref{fig:result}(b).
In the periodic limit $\rho^{SC}=1$, of course, we get $\rho^N_*=1$. In addition, we explore the disorder effect in the presence of the random chemical potentials $\mu_i\in\mu(1-\xi,1+\xi)$. For $\xi\le0.05$ level of the disorder, we observe the same power-law behavior in Eq.\eqref{mainresult} with $\beta\approx3.99$. 
%\tcg{Also, for low $\xi$, the power $\beta$ remains around the value for the case without the disorder, $\beta\approx3.99$. }
See Supplementary Materials Sec.4 for detailed information. 
%\tcb{Thus, although the size of the gap itself would not be the robust quantity, the pattern-dependent proximity effect on the strength of quasi-periodicities is maintained to the random disorder.}
Although the gap size is not the robust quantity, the pattern-dependent proximity effect is robust in the presence of the small disorder effect.

Another important character is that the quasi-periodic system enhances the stability of the proximity induced Majorana edge mode, compared to the periodic system. Specifically, red squares in Fig.\ref{fig:result} (d) illustrates that the maximum gap size, $\Delta_g^M$, is bigger as the strength of the quasi-periodicity gets stronger. Hence, $\Delta_g^M$ has a minimum value when the system is in the uniformly periodic limit, $\rho^N=\rho^{SC}=1$. Such enhancement of the stability is not an accidental one but universal for general quasi-periodic tilings. To support our result, we also investigate the influence of the random disorder on $\Delta_g^M$. In this case, we consider random disorder on either chemical potentials or pairing potentials. 
It turns out that the tendency of the enhanced stability with larger $\Delta_g^M$ is robust in the presence of small disorder effect. See Supplementary Materials Sec.4 for details.

 %For example, for $M=F_n$, Fibonacci tiling, we set $t_S^{(SC/N)}=\Delta_S^{(SC/N)}=\frac{c^{(SC/N)}}{F_{n-2}+\rho^{(SC/N)} F_{n-1}}$ for a given $\rho^{(SC/N)}$.
%\tcr{before introduce TM pattern, state details the main message for Fibonacci tiling with stacking with explaning Fibonacci part in Fig.3 }
%\tcb{No. it is not so good choice. First, each paragraph should have the single message. In this section, First paragraph explains "stability should be studied". Second paragraph explains "Stacking is necessary to study the stability". Third paragraph explains "how stability is originating in the stacking wires". And Fourth paragraph+Fifth one (originally in a paragraph), "What we will look for and how". If we write result together, }

For comparison, we also take into account another tiling pattern, Thue-Morse (TM) pattern and investigate the size of the topological gap, $\Delta_g$. Similar to the Fibonacci case, the TM pattern is also composed of two types of prototiles, say $L$ and $S$. However, unlike the Fibonacci case, the TM tiling is generated by the substitution map, $L\to LS$ and $S\to SL$, which has identical number of each prototile. In this case, the number of prototile $L$ and $S$ is the same as the alternating periodic (AP) system such as $LSLSLS \cdots$.
%alternating periodic (AP) pattern, but the ordering of the prototiles i.e., tiling pattern is different. 
Thus, the comparison between TM and AP reveal the significance of the tiling pattern rather than the statistics of the prototiles and the simple strength of the quasi-periodicity. In detail, we compare $\Delta_g^M$ for the case where the trivial metallic wire has a TM pattern, and the superconductor has a TM pattern and an AP pattern, respectively. We set $M=610$.

By exploring the $\Delta_g$ in the TM pattern, we observe that the similar power-law relationship between $\rho^{SC}$ and $\rho^N_*$ but different $\beta$ value (See green circles in Fig.\ref{fig:result} (b).). For given parameters $\tau=\delta=1$, which are the same as the Fibonacci case, we get $\beta\approx3.17$ for the TM pattern case. 
In Fig.\ref{fig:result} (d), blue circles represent the maximal gap value $\Delta_g^M$ as a function of quasi-periodic strength of the normal metal, $\rho_*^N$. Compared to $\Delta_g^M $ for the AP case, we observe that $\Delta_g^M$ for the TM pattern is \textit{always larger}, i.e., $\Delta_g^M (TM) - \Delta_g^M (AP) > 0$, as shown in green triangles in Fig.\ref{fig:result} (d). Thus, the proximity Majorana edge mode is more stabilized in the TM pattern than AP case.

Comparing the two quasi-periodic systems, Fibonacci and TM patterns, different $\beta$ for each case indicates that the power $\beta$ is indeed a pattern-dependent quantity. To understand it, we again note that the stable proximity induced Majorana edge mode is originated from the robust parallelogram cells illustrated in Fig.\ref{fig:result} (c). Then, the hoppings and pairings should be ordered to balance each link to form more robust parallelogram pairing cells, In other words, if $t_{i,i+1}^{SC}<t_{j,j+1}^{SC}$, then the case of $t_{i,i+1}^{N}<t_{j,j+1}^{N}$ forms more stable proximity induced Majorana edge mode than the case of $t_{i,i+1}^{N}\ge t_{j,j+1}^{N}$. Hence the ordering of the hoppings and pairings in the normal metal is induced by the one in the superconductor. Thus, the optimization strengths of the quasi-periodicity should have pattern-dependent relationship.
More surprisingly, we clarify that the tiling pattern is indeed critical in proximity effect by comparing the AP and TM cases. Although both tiling patterns have the same numbers of the prototiles, when the trivial metallic wire admits TM tiling pattern, the Majorana edge mode transferred to the trivial metallic wire has enhanced stability.
Thus, the enhanced stability of the Majorana edge mode depends not only on the strength of quasi-periodicity but also on the tiling pattern itself.
%Note that the enhancement is smaller than the enhancement originated from the strength of the quasi-periodicity compared to the uniformly periodic case. It is because TM pattern and an AP pattern are similar to each other. Thus, their difference could be handled as a disorder, which should be relatively small.

\section{Discussion and Conclusion}
\label{sec:conclusion}
In summary, we have argued the advantage of the Majorana edge mode in the quasi-periodic quantum wire. It lies on the presence of more stable Majorana modes originated from the pattern-dependent proximity effects. We find that the proximity effect in the superconducting wire prefers the pairings to follow the same types of pattern complexity as the hopping magnitudes. More importantly, we investigate the gap size in the stacked wires, and show that it also induces enhanced stability of the Majorana mode when two adjacent wires have the same quasi-periodic tiling patterns with induced quasi-periodicity rather than uniformly periodic one. Furthermore, we observe that the strength of the induced quasi-periodicity follows unique pattern-dependent power-law relationship, $\rho^N_*=(\rho^{SC})^\beta$, where $\beta$ is a pattern-dependent quantity. Moreover, the power-law behavior and the value of $\beta$ are maintained under low level of random disorder. Based on our study, we suggest the utility of one-dimensional quasi-periodic systems and their proximity effect for a stabilized braiding of the Majorana modes.

Beyond the one dimensional quasi-periodic chains, the pattern-dependent proximity effects in the two-dimensional systems including novel heterostructures, such as twisted bilayer graphene would be also an interesting future work\cite{xu2018topological,can2021high}\footnote{Junmo Jeon and SungBin Lee, in preparation}. We also suggest that our study can be experimentally examined based on the anisotropic XY spin chains with quasi-periodic patterns.
\cite{coldea2010quantum,degottardi2011topological}.

\subsection*{Acknowledgement}
We are greateful to Haru.K. Park for useful comments. This work is supported by National Research Foundation Grant (NRF-
2020R1F1A1073870, NRF-2020R1A4A3079707,NRF grand 2021R1A2C1093060). 

\bibliography{my1}
\bibliographystyle{unsrt}

%%%%%%%Supple

\clearpage
\pagebreak

\renewcommand{\thesection}{\arabic{section}}
\setcounter{section}{0}
\renewcommand{\thefigure}{S\arabic{figure}}
\setcounter{figure}{0}
\renewcommand{\theequation}{S\arabic{equation}}
\setcounter{equation}{0}

\begin{widetext}
	\section*{Supplementary Material for Pattern-dependent proximity effect and Majorana edge mode in one-dimensional quasicrystals}

	\section{Topological invariant}
	\label{sec:1}

In this section, we derive the simplified topological invariant, $\nu$, in the main text\cite{degottardi2013majorana}. For derivation, we assume that hopping magnitudes and pairing potentials are real values. Hence the global transfer matrix, $\mathcal{A}_p$, is also real $2\times 2$ matrix that can be written as $\mathcal{A}_p= \begin{pmatrix} a & b \\ c & d \end{pmatrix}$, where $a,b,c$ and $d$ are all reals. Then, its eigenvalues are given by $\lambda_{\pm}=\frac{a+d\pm\sqrt{(a+d)^2-4(ad-bc)}}{2}$. We first note that if two eigenvalues are not real, then they are complex conjugate pair. Thus, if there is only one eigenvalue whose magnitude is less than 1 (that is the case of trivial phase), then it should be real eigenvalue, and hence both are reals. Therefore, trivial phase is given by the negative $(\lambda_+^2-1)(\lambda_-^2-1)$. However, since $(\lambda_+^2-1)(\lambda_-^2-1)=(\lambda_++1)(\lambda_-+1)(\lambda_+-1)(\lambda_--1)=\mbox{det}(\mathcal{A}_p+I)\mbox{det}(\mathcal{A}_p-I)$, we have $\nu=-\mbox{sgn}(f(1)f(-1))$, where $f(z)=\mbox{det}(\mathcal{A}_p-zI)$. Furthermore, the $f(z)$ becomes singular when $z=\lambda_{\pm}$. Hence we have $\nu=(-1)^{n-1}$, where $n=\frac{1}{2\pi i}\oint_{|z|=1} \frac{f'(z)}{f(z)}$ is topological winding number.

\section{Derivation and discussion about the topological phase under $\Delta_{i,i+1}=t_{i,i+1}$}
	\label{sec:2}

In this section, we derive the topological invariant under $\Delta_{i,i+1}=t_{i,i+1}$ limit (Eq.(4) in the main text) and discuss its influence. Eq.(4) in the main text shows that the topological invariant under $\Delta_{i,i+1}=t_{i,i+1}$ is given by
\begin{align}
\label{quasivs}
\nu=\mbox{sgn}\left(\left(\frac{\mu}{2t_S\rho^{\alpha(M)}}\right)^{2M}-1\right).
\end{align}
Here, $M$ is the length of the system. See the main text for detailed definitions of the variables. To derive Eq.\eqref{quasivs}, we first note that the second column for each transfer matrix in Eq.(3b) in the main text becomes zero under $\Delta_{i,i+1}=t_{i,i+1}$. Also, we note that the set of matrices with zero second column forms a monoid whose identity element is given by $\begin{pmatrix} 1&0 \\ 0&0 \end{pmatrix}$. Thus, the global transfer matrix, $\mathcal{A}_p$ and its square also have zero second column. Then, when we let $a$ for (1,1) component of $\mathcal{A}_p^2$, the topological invariant is given by $\nu=\mbox{sgn}(a-1)$. Since the (1,1) component of each transfer matrix in Eq.(3b) in the main text is $\frac{\mu}{2t_{i,i+1}}$, we can achieve $a=\left(\prod_{i=1}^N\frac{\mu}{2t_{i,i+1}}\right)^2$ from the mathmatical induction. Note that the chemical potential is uniform in our paper. Let us assume that there are $M\alpha(M)$ number of $L$ tiles among  $M$ number of tiles. Then, $a=\left(\frac{\mu}{2t_S\rho^{\alpha(M)}}\right)^{2N}$, and hence we get Eq.\eqref{quasivs}.

Next, let us discuss the influence of Eq.\eqref{quasivs}. By exemplifying Fibonacci tiling, we survey the energy spectrums and the size of the topological gap for three different strengths of the quasi-periodicity, $\rho=0.5,1$ and $2$, where $\rho\equiv t_L/t_S$. Note that we are taking into account $\Delta_{i,i+1}=t_{i,i+1}$ limit. Especially, $\rho=1$ represents the uniformly periodic case. For comparison, we keep the average pairings, and hence the average hoppings as the main text i.e. $\sum_{i=1}^{M}t_{i,i+1}=cM$, where $c$ is a constant. Then, from Eq.\eqref{quasivs}, the critical chemical potential, $\mu_*$, which divides the topological phase and trivial phase, depends on the strength of the quasi-periodicity. Specifically, the case of $\rho>1 (<1)$ establishes $\mu_*>2t_S (<2t_S)$, while $\mu_*=2t_S$ for $\rho=1$ (See Fig.\ref{figsup} (a-c)). Furthermore, we investigate the size of the gap, $\Delta_g$ as a function of $\mu/2t_S$. As a result, we discover that $\Delta_g$ is a curved function of $\mu/2t_S$ when the system admits the nontrivial quasi-periodicity (See Fig.\ref{figsup} (d).), on the other hand, $\Delta_g$ is a linear function of $\mu/2t_S$ for the periodic limit. Note that $\Delta_g$ is proportional to $1-\frac{\mu}{2t_S\rho^{\alpha(M)}}$. Hence the nontrivial quasi-periodicity with $\alpha(M)\neq0$, we have curved behavior of $\Delta_g$ in quasicrystal as contrast to the uniformly periodic system.
\begin{figure}[]
	\centering
	\includegraphics[width=0.7\textwidth]{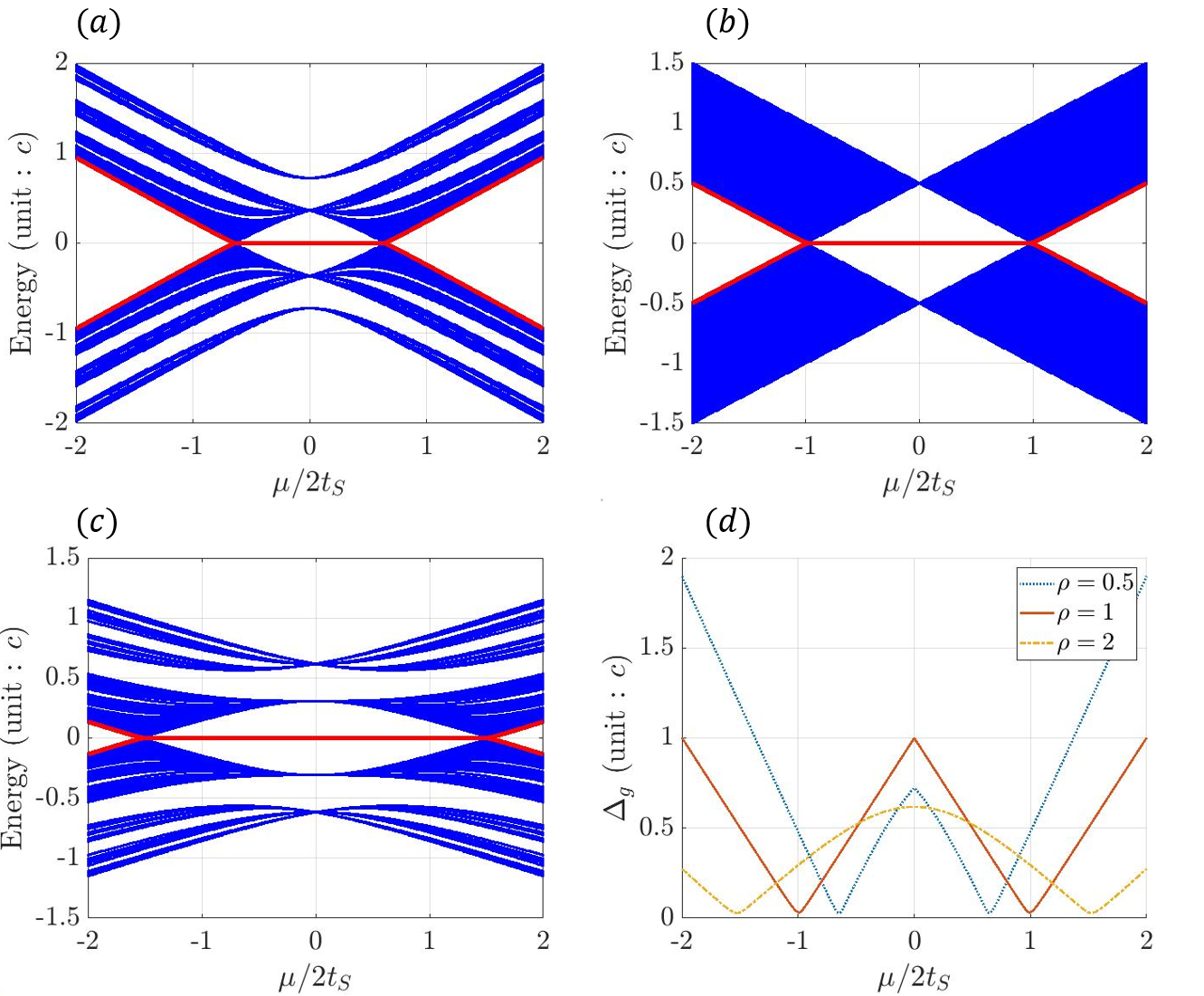}
	\label{figsupf}
	
	\caption{\label{figsup}(a-c) Energy spectrums of Kitaev chain model as a functions of $\mu/2t_S$ on Fibonacci quasicrystal with three different strength of the quasi-periodicity (a) $\rho=0.5$ (b) $\rho=1$, periodic limit, and (c) $\rho=2$, respectively. Red lines represent the middle states. The Majorana edge modes emerge when their is no gap in between middle states. (d) The size of the topological gap, $\Delta_g$ as a function of $\mu/2t_S$. The periodic limit (orange line) shows a linear behavior. In contrast, the quasi-periodicity induces the curvature on the relation (yellow and blue dashed lines). See the main text for details.
	}
\end{figure}

\section{The origin of the proximity induced Majorana edge mode of stacked quantum wires}
	\label{sec:3}
In this section, we discuss the general origin of the emergent proximity induced Majorana edge mode of stacked quantum wires (See Fig. 1 in the main text). Originally, one wire is the topological superconductor, while the other one is trivial normal metal. However, as turning on the inter-wires hoppings and pairings, the proximity induced $b$-Majorana edge mode emerges on the right end of the normal metal. Since probability is conserved, there should be Majorana modes in superconducting wire, which have complement probability of it. Given $\rho^{SC}=0.8$ Fibonacci pattern on the superconducting wire, Fig.\ref{sup} (a) illustrates that the probabilities of the $b$-Majorana edge modes on the right end of the superconductor ($P(b^{SC}_{M+1})$, blue color) and normal metal ($P(b^{N}_{M+1})$, orange color) as the functions of the strength of inter-wires pairing, $\delta$, respectively. It shows that the Majorana edge mode on the same end of the superconductor is always the origin of the proximity induced Majorana edge mode. We also compute the two-site correlation function, $\left\langle ib_{M+1}^{SC}b_{M+1}^{N}\right\rangle$, in Fig.\ref{sup} (b). The inter-wires pairing, $\delta$, enhances the correlation function when it is small. On the other hand, it reduces the correlation function for sufficiently large inter-wires pairing regime. It is because the strong inter-wires coupling merges two wires as a single one.
\begin{figure}[]
	\centering
	\includegraphics[width=0.7\textwidth]{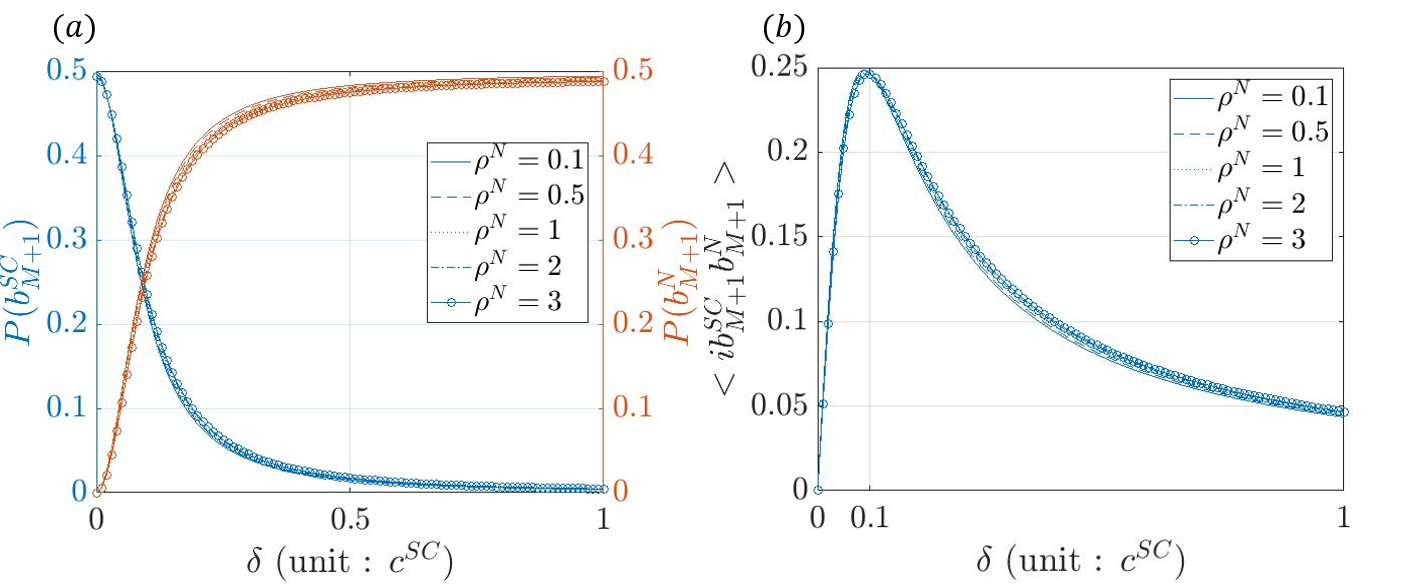}
	\label{figsupp}
	
	\caption{\label{sup}Demonstration of the proximity induced Majorana edge mode. Here, we fix $\rho^{SC}=0.8$. (a) The probability of the $b$-Majorana fermion on the right end of each wire as a function of the inter-wires pairing strength, $\delta$. Regardless of the strength of quasi-periodicity, $\rho^N$, two probabilities are complementary. (b) The two-site correlation function, $\left\langle ib^{SC}_{M+1}b^{N}_{M+1}\right\rangle$ as a function of $\delta$. The small $\delta$ enhances the two-site correlation function, but sufficiently large $\delta$ rather reduces it. See the main text for the detailed information.
	}
\end{figure}

\section{Robustness of the pattern-dependent proximity effects under the disorder}
	\label{sec:4}
In this section, we discuss the random disorders in the stacked quantum wires and their influence on the pattern-dependent proximity effect.  %\tcg{that is given by the power-law relationship, $\rho_{*}^N=(\rho^{SC})^\beta$ and the enhanced stability of the proximity induced Majorana edge mode by the quasi-periodicity}.
Note that the results are illustrated in Fig.2 (b) and (d) in the main text. 
%\tcr{explain Fig.2 b and d}
%\tcg{The disorder emerges on the chemical potentials as randomly $\mu_i\in\mu(1-\xi,1+\xi)$, where $\mu_i$ is the chemical potential of the $i$-th site. }

The random disorders are given by either $\mu_i\in\mu(1-\xi,1+\xi)$ or $\Delta_{i,i+1}\in\Delta_{L/S}(1-\zeta,1+\zeta)$, where $L,S$ are the type of the link $(i,i+1)$ and $\Delta_{L/S}$ are the variational pairings for each link, respectively. Here, $\xi$ and $\zeta$ indicate the levels of the random disorder.

%\tcb{Similarly, due to the self-consistency, the pairing potentials would have the disorder as $\Delta_{i,i+1}\in\Delta_{L/S}(1-\zeta,1+\zeta)$, where $L,S$ are the type of the link $(i,i+1)$ and $\Delta_{L/S}$ are the variational pairings for each link, respectively. Here, $\xi$ and $\zeta$ represent the levels of the disorder. For instance, $\xi=0.01$ corresponds to the $1\%$ level of the disorder.}
%\tcr{due to the self-consistency???? clarify the sentence} 

We investigate $\rho_{*}^N$ as a function of $\rho^{SC}$ for each level of random disorder on the chemical potential, $\xi$ for the Fibonacci pattern.
By using the linear regression, we extract the tendency of $\rho_{*}^N$ as a function of $\rho^{SC}$. In detail, we illustrate Fig.2 (b) in the main text with the disorders, particularly, $\xi=0.05$ in Fig.\ref{fig:disorderC} (a). From Fig.\ref{fig:disorderC} (a), the power-law relationship is maintained under the small random disorder. Note that the general power-law relationship is given by $\rho^{N}_*=C(\rho^{SC})^\beta$. Here, $\beta$ and $C$ are determined by the slope and the $y$-intercept of the linera regression for Fig.\ref{fig:disorderC} (a), respectively. In Fig.\ref{fig:disorderC} (b), $C$ is maintained as 1 (red squares, right $y$-axis) and $\beta$ shows stationary behavior when $\xi$ is small (blue circles, left $y$-axis). Thus, despite the presence of disorder would break the tiling pattern, the power-law relationship between $\rho^{N}_*$ and $\rho^{SC}$ is maintained for the low level disorders. 
%\tcg{Note that generally, $\rho^{N}_*=C(\rho^{SC})^\beta$, where $\beta$ is the slope and $C$ is the $y$-intercept of $\log(\rho^{N}_*) \ vs \ \log(\rho^{SC})$ plot. We use the Fibonacci pattern and other parameters are all the same as the main text.} The results are illustrated in Fig.\ref{fig:disorderC} \tcr{for the Fibonacci pattern}. \tcb{Surprisingly, we could point out two messages. First, although the random disorder would break the tiling pattern, when its level is small enough, the power-law relationship is still observed (See Fig.\ref{fig:disorderC} (a).). Second, furthermore, the $\beta$ and $C$ exhibits stationary behavior for a relatively low level of disorder. Explicitly, $C$ is maintained as $1$ (red squares, right $y$-axis) and $\beta$ (blue circles, left $y$-axis) is distinguishable to the Thue-Morse (TM) pattern case (See the green dashed line in Fig.\ref{fig:disorderC} (b).). Hence, our results are stable even the random disorder effect emerges as low level.}
%\tcr{explain Fig in detail and summarize the result}
\begin{figure}[]
\centering
\includegraphics[width=0.9\textwidth]{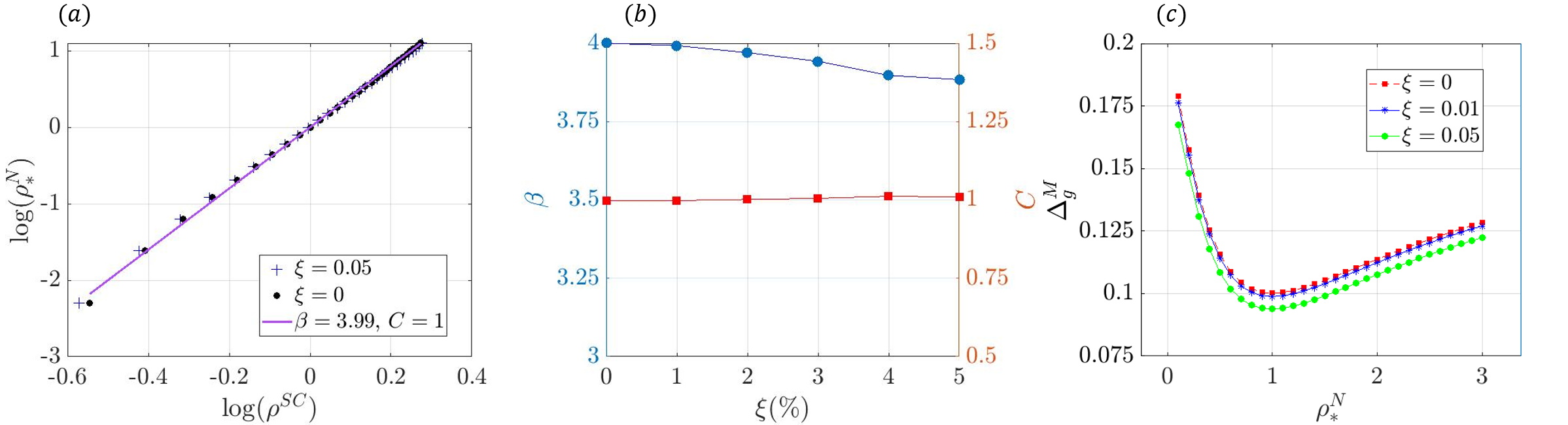}
\label{figsuppp}

\caption{\label{fig:disorderC} (a) The power-law behavior of $\rho_{*}^N$ as function of $\rho^{SC}$ under the random disorder. (b) In the Fibonacci pattern, the power-law behavior with $\beta\approx3.99$ and $C=1$ are maintained for a small $\xi$ level of disorder. Blue circles represent $\beta$ as the left $y$-axis. Red squares represent $C$ as the right $y$-axis. (c) The maximal topological gap size, $\Delta_g^M$ as a function of the strength of the quasi-periodicity encoded on the hoppings and pairings under the presence of the random disorder. As the level of disorder increases, $\Delta_g^M$ decreases for each strength of the quasi-periodicities. For any $\xi$, the quasi-periodic hoppings and pairings significantly enhance $\Delta_g^M$ compared to the disorder effect.
}
\end{figure}

Next, we survey the influence of the random chemical potentials on the maximal size of the topological gap, $\Delta_g^M$. %\tcb{As a result, even though the size of the gap would be disturbed by the strong random disorder effect, the quasi-periodicity of hoppings and pairings still enhances the stability compared to the uniformly periodic system.}
It turns out that despite the presence of disorder, the the quasi-periodic hoppings and pairings enhance the stability of the Majorana modes compared to the uniformly periodic case, as shown in Fig.\ref{fig:disorderC} (c) particularly for $\xi=0, 0.01, 0.05$.

Lastly, we ask the robustness of the gap size under random disturbance on the pairing potentials. To illustrate its robustness, we draw Fig.2 (d) for each level of disturbance, $\zeta\le0.05$ (See Fig.\ref{fig:disdis}). Interestingly, the disturbance does not
significantly change the enhancement of the gap size. Thus, we conclude that in the presence of the disturbance on the pairings, the  proximity-induced Marjorana mode has the enhanced stability in the quasi-periodic system.

\begin{figure}[]
\centering
\includegraphics[width=0.5\textwidth]{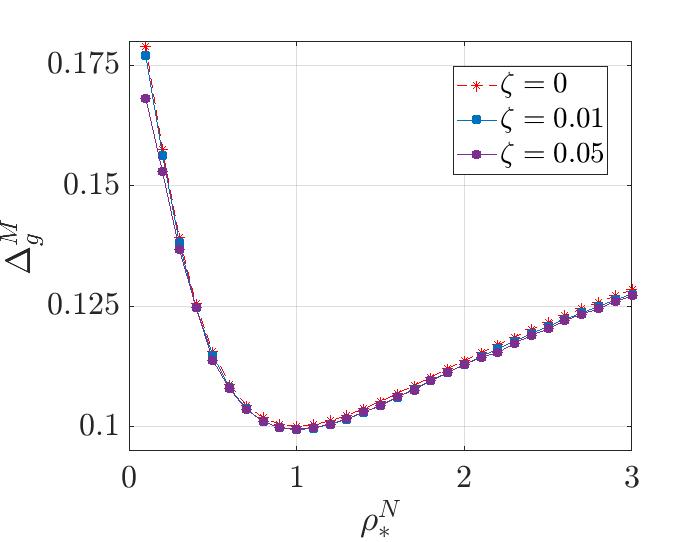}
\label{figsuppp}

\caption{\label{fig:disdis} The maximal size of the gap is robust to the local disturbance of self-consistent pairing potential. The level of local disturbance is defined by $\zeta\le0.05$. Red ($\zeta=0$, without local disturbance), blue square ($\zeta=0.01$) and violet circle ($\zeta=0.05$), respectively.
}
\end{figure}

\end{widetext}

\end{document}